\documentclass[runningheads]{llncs}
\usepackage[T1]{fontenc}
\usepackage{graphicx,verbatim}
\usepackage{booktabs}
\usepackage{multirow} 
\usepackage{tabularx} 
\usepackage{graphicx} 
\usepackage{makecell} 
\usepackage{booktabs} 
\usepackage{adjustbox} 
\usepackage{xcolor}

\title{Destroy Me: Automatic Artifact Generation for Histopathology Images}
\author{Zuzanna Krawczyk-Borysiak\inst{1, 2}\orcidID{0000-0002-3897-1078}
\email{zuzanna.krawczyk@pw.edu.pl} \and
Adam Krawczyk\inst{3, 4}\orcidID{0009-0002-4851-6789}
\email{adam.krawczyk@cs.put.poznan.pl} \and
Mateusz Miller\inst{1}\orcidID{0009-0001-4362-7271}
\and
Gabriela Kaczmarek\inst{1, 2}\orcidID{0009-0004-3650-1875}
\and
Sławomir Pakuło\inst{5}\orcidID{0000-0001-6160-9651}
\and
Małgorzata Sokół\inst{1}\orcidID{0009-0008-1483-4379}
\and
Żaneta Swiderska-Chadaj\inst{1, 2}\orcidID{0000-0003-2115-9822}
}
\authorrunning{Z. Krawczyk-Borysiak et al.}
\institute{IDEAS Research Institute, Poland \and
Warsaw University of Technology, Warsaw, Poland \and
Poznan University of Technology, Poznan, Poland \and
AKCES NCBR, Warsaw, Poland \and
Tumor Pathology Department, Maria Sklodowska-Curie National Research Institute of Oncology, Poland \\
}

\begin{document}
\maketitle             
\begin{abstract}

Deep learning’s diagnostic utility in pathology is constrained by model vulnerability to real-world data imperfections. While current strategies favor "perfect data" by filtering low-quality regions, which can lead to the loss of valuable diagnostic context, we propose a paradigm shift: engineering models to thrive in imperfect environments using "Destroy Me", a hybrid framework for realistic artifact synthesis and robust data augmentation. Our approach combines Stable Diffusion, fine-tuned to preserve morphological continuity by realistically integrating artifacts with the underlying tissue architecture, with physics-based procedural modeling to synthesize six common artifact types: tissue folds, precipitates, blur, stitching errors, dust, and pen markers. Artifact fidelity is assessed using Kernel Inception Distance (KID) and color Wasserstein distance metrics. Validating this strategy on lung adenocarcinoma pattern classification with an nnU-Net, we confirm that models trained on "destroyed" patches consistently outperform baselines on independent real-world datasets. Specifically, we observed a 10.5\% relative improvement in macro F1-score and a 15\% relative increase in the Cohen’s Kappa ($\kappa$) coefficient. Crucially, our results demonstrate that selective, impact-weighted augmentation is vital for balancing practical robustness with the preservation of subtle diagnostic features. Our method is available at GitHubLinkWillBeHere.


\end{abstract}

\section{Introduction}
The rapid advancement of computational pathology, driven by Artificial Intelligence (AI) and particularly deep learning (DL), promises a transformative era in diagnostic support. Dedicated algorithms are increasingly used for critical tasks like disease detection and classification
\cite{chen2024towards,fuchs2011computational,song2023artificial}. However, a significant barrier to their widespread clinical adoption persists: \textbf{data quality}. Routine clinical Whole Slide Images (WSIs), especially Hematoxylin and Eosin (H\&E) stained, are frequently compromised by artifacts (e.g., tissue folds, pen markers, scanning glitches)~\cite{artifacts_review}. This discrepancy creates a "domain shift", causing models trained on pristine research datasets to often fail on imperfect clinical images, leading to unreliable inference~\cite{Schomig-Markiefka2021}.

Current strategies predominantly rely on automated quality control (QC) tools~\cite{janowczyk2019,PathProfiler,TCNN,GrandQC,MoE_DCNN,Sem_Seg,WSIsegQC} to filter corrupted regions. While these tools improve data purity, this approach treats artifacts as noise to be eliminated, often discarding valuable diagnostic context. Prior research has focused on the generation of synthetic artifacts to quantify model failure modes via stress testing \cite{Schomig-Markiefka2021,WANG202154} or to improve artifact detection within QC pipelines \cite{Jurgas2024}. However, these methodologies are primarily centered on diagnostic evaluation and filtering rather than performance recovery. Moreover, subtle or complex artifacts, difficult to detect automatically, still degrade model performance, highlighting a fundamental limitation in current robustness paradigms.

\begin{figure}
    \centering
    \includegraphics[width=0.99\linewidth]{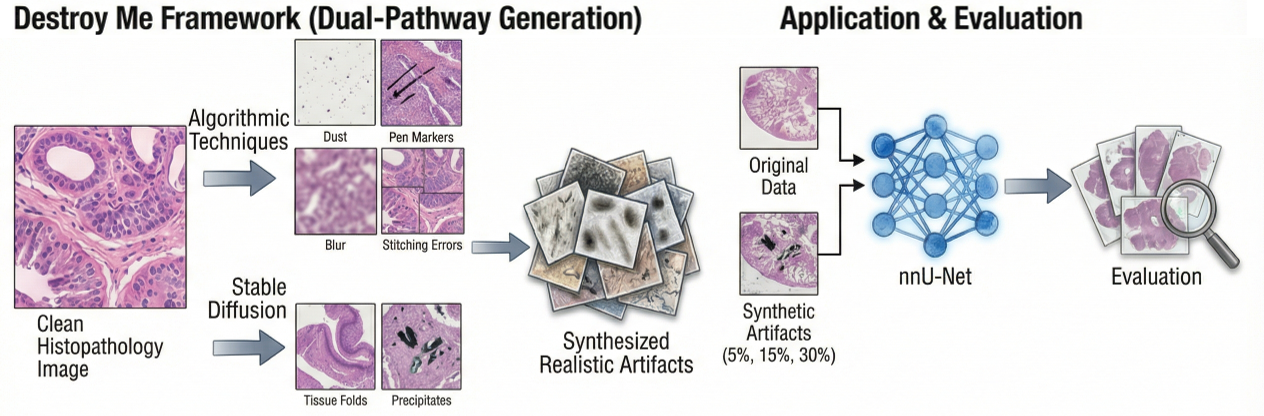}
    \caption{Overview of the proposed "Destroy Me" framework.}
    \label{fig:intro}
\end{figure}

In this work, we propose a paradigm shift: instead of striving for perfect data by eliminating artifacts, we engineer models to thrive in non-perfect real-world environments. \textbf{We introduce "Destroy Me", a novel hybrid framework that generates realistic histological artifacts for H\&E-stained tissue} (see fig.\ref{fig:intro}). Its application is primarily demonstrated and validated on lung tissue for this study.

Our approach leverages Generative AI (Stable Diffusion), fine-tuned on expert-annotated TCGA-LUAD~\cite{TCGA} data, to synthesize six common artifact types. This includes complex, texture-dependent artifacts (tissue folds, precipitates) combined with targeted algorithmic techniques for simpler degradations (blur, stitching errors, pen marker, dust).
The high perceptual realism of these generated artifacts is quantitatively confirmed by metrics like Kernel Inception Distance (KID) \cite{kid} and CLIP \cite{clip} score, showing consistent quality across diverse datasets, which underscores the method's robustness to unseen data distributions. Importantly, our research utilizes publicly available datasets, ensuring full transparency and reproducibility. The key contributions of our study are:
\begin{itemize}
    \item \textbf{Novel "Destroy Me" GenAI Framework}: A hybrid GenAI (Stable Diffusion + algorithmic) framework uniquely synthesizing six diverse, high-fidelity H\&E histological artifacts (e.g., complex folds, blur), meticulously preserving tissue morphology.
    \item \textbf{Robustness Engineering via Augmentation} We demonstrate that synthetic "destroyed" data serves as a superior augmentation strategy, enhancing model robustness to clinical noise.
    \item \textbf{Mitigating Clinical Domain Shift}: Validation on LUAD (WSI-level task) confirms that models trained with our "Destroy Me" strategy achieve superior generalization on noisy real-world clinical slides, directly mitigating the pervasive domain shift.
\end{itemize}

We apply GenAI artifact synthesis beyond mere data expansion to build robust models for imperfect pathology images, bridging the gap between research benchmarks and clinical reality.

\section{Materials and Methods}

To ensure reproducibility and transparency, this study utilized three public H\&E-stained datasets. TCGA-LUAD \cite{TCGA} was used exclusively for developing the "Destroy Me" generative framework. ANORAK \cite{anorak} served as the primary dataset for cross-validation training  of the lung adenocarcinoma (LUAD) pattern segmentation models. Finally, DHMC \cite{dhmc} provided an independent, real-world cohort for WSI-level benchmarking.

For GenAI development, subset of 73 TCGA-LUAD WSIs were manually annotated by a histotechnologist for four artifact types (dust, precipitates, folds, pen markers). We extracted 3,884 patches (384x384 px, 0.9 $\mu$m/px) with variable artifact coverage (5-95\% of patch area, and following artifact's class distribution: $57.44$\% tissue folds, $33.34$\% marker, $5.23$\% precipitates, $3.99$\% dust) and 1,131 artifact-free reference patches. Due to the technical nature of blur and stitching artifacts, which are easily reproduced procedurally, and the associated time constraints, we opted not to include these categories in the manual annotation set. Instead, the fidelity of these procedural generations was qualitatively validated through visual inspection by experienced histotechnologist. For the classification task, 731 ANORAK tiles (2000x2000 pixels) were tiled into 3,662 patches following the resolution of 384x384 px, 0.9 $\mu$m/px and its original methodology~\cite{anorak_github}. The DHMC dataset (143 WSIs) was used for external validation; these slides were tiled into patches for inference, with slide-level labels assigned based on the highest cumulative pixel count of segmented patterns across the entire WSI.

\subsection{Learning-Based Artifact Synthesis}

Tissue folds and precipitates introduce complex, occlusive textures requiring deep generative modeling to maintain structural context. Unlike static procedural overlays, our Stable Diffusion approach maintains morphological continuity by ensuring that generated artifacts realistically "wrap" around and integrate with the underlying histological architecture. We utilized a Stable Diffusion 1.5 Inpainting backbone~\cite{Rombach_2022_CVPR}, which employs a 9-channel input (noisy latents, binary mask, and masked image latents) to ensure seamless integration with the surrounding tissue.

\textbf{Training and Inference Protocol}: Models were trained on $384 \times 384$ pixel patches (batch size: 4, LR: $1 \times 10^{-5}$, FP16 precision) using the Diffusers~\cite{diffusers} library on Nvidia A100 (40GB). We used an 80:20 stratified split (by artifact type and clean tissue) for generative development. For synthesis, we employed a "Mask Bank" strategy, stochastically sampling realistic geometric footprints from real-world clinical samples. Final outputs were refined via grid search over Classifier-Free Guidance scales [5.0, 7.5, 9.0] and denoising strengths [0.8, 0.9, 1.0], with optimal values at 7.5 and 0.9, respectively. Performance was primarily evaluated using masked Kernel Inception Distance (KID) to account for limited test split sizes ($N < 500$), alongside Wasserstein distance and CLIP scores. 

\textbf{Fold}  artifacts, tissue folds that are dense, hyper-stained structural occlusions, were modeled using a dual-objective strategy. The model was trained to reconstruct ground-truth folds while simultaneously learning to inpaint healthy tissue on clean samples using random masks to prevent overfitting. During inference, Gaussian blurring ($\sigma=3.0$) was applied to mask edges to ensure textural continuity between the generated crease and the underlying tissue.

\textbf{Precipitates}, representing localized dye aggregations or external contaminants with diverse geometric profiles were synthesized using a hybrid training set, augmenting sparse real-world data with procedural samples from Section \ref{section:procedural_alg}. To allow for organic crystal growth beyond the original mask boundaries, masks underwent max-pool dilation (kernel 21) prior to inference. Text-conditioning emphasizing "semi-transparent blending" was utilized to ensure the underlying tissue structure remained visible through the generated deposits.

\subsection{Procedural and Physics-Based Synthesis} \label{section:procedural_alg} Stitching and blurring artifacts, characteristic of the digitization phase, are deterministic and arise from well-defined mechanical or optical aberrations. Unlike the complex, stochastic textures of tissue folds or chemical precipitates, these artifacts can be simulated with greater precision, controllability, and computational efficiency using explicit image processing operations rather than data-hungry generative models.

\textbf{Stitching} artifacts occur when the scanner's software fails to align adjacent high-resolution tiles, resulting in visible discontinuities. To simulate this, we define a grid within the image and apply a random displacement ($\Delta x, \Delta y$) to adjacent regions. To maintain image continuity and avoid zero-padding artifacts, we employ reflection padding on the source image prior to extraction.

\textbf{Blur} artifacts, typically caused by focus plane errors are simulated by applying a Gaussian blur kernel to specific regions of interest. To ensure realistic, gradual transitions between sharp and blurry areas, mimicking the physics of an optical lens, we apply a feathered alpha-masking technique to blend the blurred region with the original tissue.

\textbf{Dust}, representing physical debris introduced during slide preparation or scanning, is modeled as a spatial intensity map. This map is constructed by stochastically distributing overlapping elliptical sub-components along a principal axis with localized jitter to simulate directionally smeared, organic structures. The model incorporates an optional dense elliptical core and is modulated by a grit noise map (generated via a normal distribution) to replicate particulate texture. The local opacity of the artifact is defined by:
\begin{equation}
\label{eq:dust_alpha_mask}
\alpha(p) = |(\sum P_i(p) + C(p)) \cdot N(p)| * G_{\sigma}
\end{equation}
where $p$ is pixel, $P_i(p)$ are stochastically distributed elliptical "puffs", $C(p)$ is the dense elliptical core, $N(p)$ is the grit noise map, and $G_{\sigma}$ is a Gaussian defocus blur. To simulate the narrow depth of field in optical microscopy, a conditional defocusing mechanism is applied to the input tissue $I_{in}$: if the artifact coverage exceeds a spatial threshold (0.15), the underlying tissue is subjected to a heavy Gaussian blur to create $I_{defocus}$. Light scattering is subsequently simulated by alpha-blending the composite intensity map with the (potentially defocused) tissue:
\begin{equation}
\label{eq:dust_alpha_blending}
I_{out} = I_{defocus} \odot (1 - \alpha) + C_{dust} \odot \alpha
\end{equation}
where $\odot$ is Hadamard Product, $C_{dust}$ represents the normalized color vector of the contaminant.

\textbf{Marker}
artifacts are common histopathological contaminant occurring when pathologists annotate slides; while these marks are often translucent, they can obscure underlying tissue architecture. We model this artifact using a two-stage procedural approach. First, the spatial footprint of the marker is generated by anchoring random polygons to the image boundaries, simulating the large-scale annotations typically observed at high magnification. The non-uniform density of the ink, which is characteristically more saturated in the center of the stroke, is modeled using a power-law transformed Euclidean distance transform (EDT). The local opacity of each pixel is defined by:
\begin{equation}
\label{eq:marker_alpha_mask}
\alpha(p) = \left( \Phi(p) \right)^{\gamma} \cdot (M * G_{\sigma})(p) \cdot \alpha_{max}
\end{equation}
where $\Phi(p)$ is the normalized Euclidean distance from pixel $p$ to the nearest boundary of the binary mask $M$, $\gamma$ is the density exponent (e.g., $0.35$ for heavy ink, $0.70$ for standard), $G_{\sigma}$ is a Gaussian kernel representing the capillary "bleed" at the mask edges, $\alpha_{max}$ is the peak opacity constant.
To ensure physical realism and prevent artificial tissue bleaching, we employ a subtractive blending model to simulate light attenuation through the pigment:
\begin{equation}
\label{eq:marker_alpha_blending}
I_{out} = I_{in} \odot [ 1 - \alpha \odot (1 - C_{ink}) ]
\end{equation}
where $I_{in}$ and $I_{out}$ represent the input and resulting pixel intensities, respectively, and $C_{ink}$ denotes the normalized pigment color vector.

\textbf{Precipitates} were modeled by stochastically generated four spatial mask types: crystalline structures (convex hull of $n$ points), fluid dye pools (overlapping Gaussian-blurred ellipses), fibrous threads (thick, striated fibers), and granular specks. These subtypes vary in opacity from semi-translucent fibers to dense, nearly opaque crystals. To simulate optical refraction through crystalline structures, we apply a coordinate displacement mapping:
\begin{equation} \label{eq:refraction}
I_{refr} = I_{in}(p + \Delta p)
\end{equation}
where $\Delta p = (\Delta x, \Delta y)$ is a random displacement vector applied within the crystalline mask boundaries. We treat precipitates as light-absorbing filters and model the final intensity $I_{out}$ using the subtractive blend defined in Eq. \ref{eq:marker_alpha_blending}, substituting $I_{in}$ with $I_{refr}$ and using a gritty texture map $C_{tex}$ (blended base stain color and Gaussian noise) in place of $C_{ink}$. These procedural samples were then used to augment our training set for the Stable Diffusion model, enabling the GenAI to learn complex textural variations and organic blending patterns that exceed simple algorithmic modeling.

\section{Results}

To evaluate the impact of the "Destroy Me" framework, we compared a baseline nnU-Net~\cite{Ise_nnUNet_MICCAI2024} against models trained with varying levels of artifact augmentation. For individual artifacts, we utilized in-place augmentation, where 5\%, 15\%, or 30\% of the training patches were replaced by their synthetic-artifact versions to maintain a constant dataset volume. Multi-artifact configurations (Mix 1–3) utilized a combination of in-place replacement (Mix 1) and dataset expansion (Mix 1-2) to test the limits of model generalization. We prioritize the macro-F1 score to account for class imbalance, with Cohen's Kappa ($\kappa$) providing context on inter-observer agreement.

\begin{figure}
    \centering \includegraphics[width=0.85\linewidth]{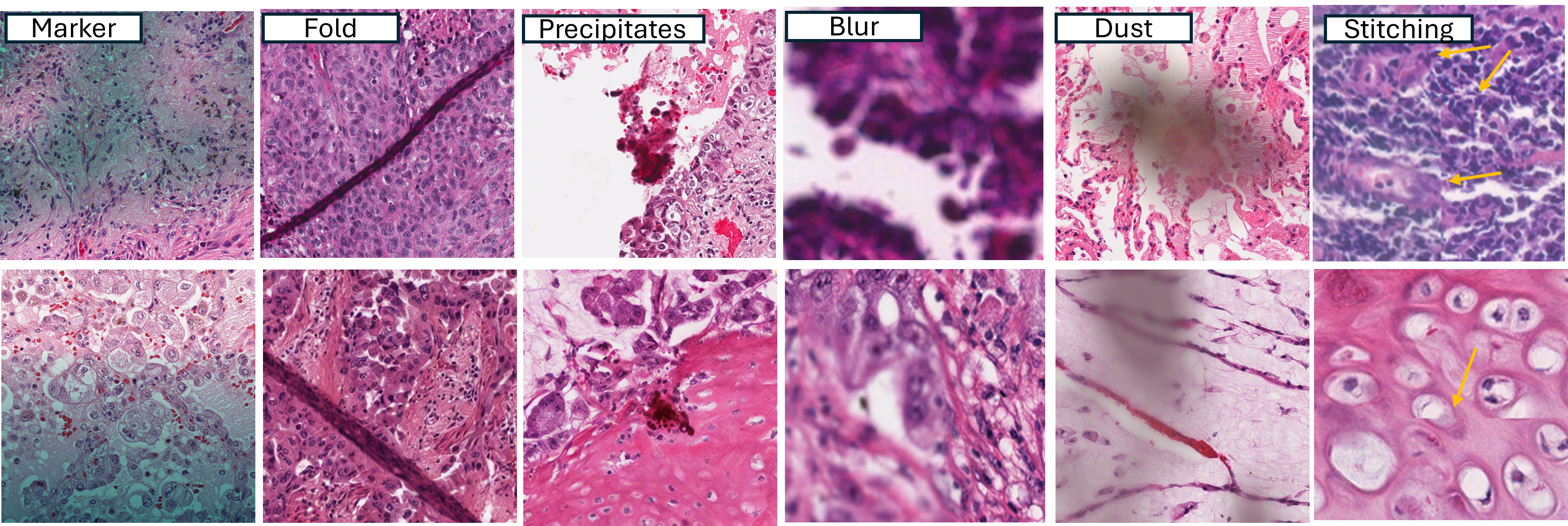}
    \caption{Examples of real (row 1) and generated (row 2) artifacts.}
    \label{fig:patches}
\end{figure}

Figure \ref{fig:patches} presents examples of patches with real and generated artifacts, whereas Table \ref{tab:results_general_merged_final} summarizes the overall performance. The baseline model, trained solely on clean data, achieved an macro-F1 of 0.38 and $\kappa$ of 0.40, corresponding to fair agreement. Incorporating synthetic artifacts via the "Destroy Me" framework consistently improved performance across all categories. The most substantial gains were observed with generative folds and procedural markers, both at 30\% in-place augmentation, yielding the highest overall scores (macro-F1: 0.42, $\kappa$: 0.46 -- moderate agreement). This represents a relative improvement of 10.5\% in macro-F1 and 15\% in $\kappa$ over the baseline. Simpler algorithmic augmentations (stitching, blur) and generative precipitates peaked at a 15\% ratio (macro-F1: 0.40–0.41), while dust augmentation showed a steady upward trend, reaching an macro-F1 of 0.42 at 30\% intensity. 

We evaluated multi-artifact training through three configurations: Mix 1 (static volume; $30\%$ in place augmentation by artifacts at $5\%$ per category); Mix 2 (expanded by $135\%$; $30\%$ each for markers, folds, and dust and $15\%$ for others); and Mix 3 (expanded by $75\%$; $20\%$ each for markers, folds, and dust, and $5\%$ for others). While both expanded sets reached a performance plateau with a $10.5\%$ improvement in macro-F1 and a $15\%$ increase in Cohen’s $\kappa$, Mix 3 emerged as the superior solution. Despite its lower data volume, Mix 3 demonstrated greater categorical stability across all LUAD grade patterns, whereas the more aggressive Mix 2 expansion resulted in a performance degradation specifically within the papillary pattern. This indicates that excessive data expansion can introduce over-regularization by causing the model to misidentify subtle diagnostic features as artifacts, which highlights why selective and impact weighted augmentation provides the most efficient path to practical robustness.

\begin{table}[h!]
\centering
\fontsize{8pt}{9.6pt}\selectfont 
\setlength{\tabcolsep}{1.2pt} 
\caption{
Comparison of "Destroy Me" augmentation strategies vs. baseline. Results: macro-F1/Kappa ($\kappa$). Best result per ratio bolded. \textbf{Imp.(\%)}: Relative procentage improvement over baseline for macro-F1.}
\label{tab:results_general_merged_final}
\begin{tabular}{@{}l|c|llllll|c@{}} 
\toprule
\textbf{Ratio} & \textbf{Baseline} & \textbf{Stitching} & \textbf{Blur} & \textbf{Dust} & \textbf{Folds} & \textbf{Precip.} & \textbf{Marker} & \textbf{Imp.} \\
\midrule
\textbf{0\%}   & 0.38/0.40 & -- & -- & -- & -- & -- & -- & -- \\ 
\midrule 
\textbf{5\%}   & -- & 0.39/0.41 & 0.40/0.42 & 0.40/0.42 & 0.40/0.44 & 0.40/0.43 & 0.41/0.44 & 7.9\% \\ 
\textbf{15\%}  & -- & \textbf{0.40/0.43} & \textbf{0.41/0.43} & 0.40/0.43 & 0.41/0.43 & \textbf{0.41/0.44} & 0.41/0.43 & 7.9\% \\ 
\textbf{30\%}  & -- & 0.39/0.41 & 0.40/0.43 & \textbf{0.42/0.45} & \textbf{0.42/0.46} & 0.40/0.43 & \textbf{0.42/0.46} & 10.5\% \\ 
\midrule 
\multicolumn{9}{@{}c}{Mixed Artifacts} \\ 
\midrule 
\textbf{Mix 1}& 0.38/0.40& \multicolumn{6}{c|}{0.40/0.42} & 5.3\%  \\
\textbf{Mix 2}& 0.38/0.40& \multicolumn{6}{c|}{0.42/0.46} & 10.5\%  \\
\textbf{Mix 3}& 0.38/0.40 & \multicolumn{6}{c|}{0.42/0.46}& 10.5\% \\
\bottomrule
\end{tabular}
\end{table}

\subsection{Quantitative Assessment of Artifact Fidelity}

We assessed artifact realism using KID \cite{kid}, CLIP \cite{clip}, 1D Wasserstein distance \cite{stain_normalisation}, and Masked SSIM to quantify structural preservation. KID was prioritized over FID for its unbiased estimation in small histopathological subsets \cite{Yash_metrics}.

As shown in Table \ref{tab:fidelity_metrics}, our framework achieves high-fidelity synthesis across domains. Internal TCGA-LUAD KID values (0.0197--0.0444) outperform established thresholds (0.04--0.06) for medical GenAI \cite{PathDiff,OhJin_Pathologyaware_MICCAI2025}. CLIP scores (30.80--32.95) confirm accurate zero-shot semantic grounding \cite{clip}. Notably, Masked SSIM values distinguish the artifacts' physical nature: high scores for dust and markers ($\sim$0.73--0.82) confirm the preservation of underlying cellular morphology, while lower scores for folds and precipitates ($\sim$0.14--0.27) correctly reflect their structurally destructive, occlusive properties. Wasserstein distances (25.98--42.70) reflect natural inter-laboratory staining heterogeneity rather than generative error \cite{stain_normalisation}. Stable KID metrics on the unseen ANORAK dataset (max 0.0536) highlight the framework's robustness to domain shift without dataset-specific fine-tuning.

\begin{table*}[ht!]
\centering
\fontsize{8pt}{9.6pt}\selectfont
\caption{Quantitative evaluation of generated artifact realism on ANORAK and TCGA-LUAD datasets. Lower KID and Color Distance ($\downarrow$) indicate higher fidelity; higher CLIP Score ($\uparrow$) means better perceptual similarity. Masked SSIM measures structural preservation (high for additive, low for occluding artifacts). N/A entries signify metrics not applicable for algorithmically generated artifacts.}
\label{tab:fidelity_metrics}
\begin{tabular}{@{}l c c c c | c c c c@{}}
\toprule
\textbf{Artifact} & \multicolumn{4}{c|}{\textbf{ANORAK Dataset}} & \multicolumn{4}{c}{\textbf{TCGA-LUAD Test Set}} \\
\cmidrule(lr){2-5} \cmidrule(lr){6-9}
& \textbf{KID ($\downarrow$)} & \textbf{CLIP ($\uparrow$)} & \textbf{Color ($\downarrow$)} & \textbf{SSIM} & \textbf{KID ($\downarrow$)} & \textbf{CLIP ($\uparrow$)} & \textbf{Color ($\downarrow$)} & \textbf{SSIM} \\
\midrule
Folds        & 0.0435 & 30.80 & 42.70 & 0.269 & 0.0257 & 30.95 & 28.67 & 0.149 \\
Precipitates & 0.0536 & 30.86 & 42.41 & 0.238 & 0.0444 & 30.85 & 28.86 & 0.140 \\
Dust         & 0.0399 & N/A   & 35.18 & 0.819 & 0.0197 & N/A   & 25.98 & 0.810 \\
Marker       & 0.0406 & N/A   & 40.33 & 0.740 & 0.0275 & N/A   & 38.43 & 0.733 \\
\bottomrule
\end{tabular}
\end{table*}

Regarding efficiency, SD-based synthesis averaged 1.42s/patch, whereas procedural methods (0.05-0.08s/patch) enabled high-throughput clinical augmentation.

\section{Discussion}
The primary finding of this study is that the proposed "Destroy Me" strategy consistently enhances model generalization on real-world clinical data. This is achieved without the need for collecting or manually annotating additional real-world artifacts, a process which is both costly and time-consuming. Our method led to a relative improvement of 10.5\% in macro-F1 (from 0.38 to 0.42) and an increase in $\kappa$ score from 0.40 to 0.46 (from fair to moderate agreement). By systematically introducing synthetic imperfections, our framework enables networks to learn robust, morphology-invariant features instead of relying on fragile visual cues present only in pristine slides. This approach strongly aligns with the principles of Data-Centric AI, where realistically simulating data imperfections yields greater returns than architectural tuning alone. By simulating realistic clinical imperfections, we democratize access to robust AI development and reduce the risk of model failure in practice, a prerequisite for clinical trust.

This work represents a crucial step towards adapting AI models for real-world clinical use and increasing their deployment possibilities. By engineering robustness against common pre-analytical variables, we reduce the risk of model failure on non-ideal clinical slides, which is critical for building trust in AI diagnostics. Our dual-engine framework, powered by advanced Generative AI (Stable Diffusion), is uniquely suited for this task. Unlike simple algorithmic noise, GenAI generates context-aware, complex artifact textures (e.g., folds) that provide a richer training signal, as confirmed by high fidelity metrics (Table \ref{tab:fidelity_metrics}) and consistent quality across diverse datasets. Beyond its role in augmentation, this "Destroy Me" framework also serves as a valuable tool for benchmarking AI models, allowing developers to precisely identify specific vulnerabilities to different artifact types and tailor their solutions accordingly.

Future work includes extending this approach to other tissue types as well as exploring multi-artifact compositions to further simulate the complexity of clinical pathology.

 \bibliographystyle{splncs04}
 \bibliography{main}

\end{document}